\documentclass{osa-article}

\journal{osac}

\articletype{Research Article}

\begin{document}

\title{Genuine time-bin-encoded quantum key distribution over a turbulent depolarizing free-space channel}

\author{Jeongwan Jin,\authormark{1,$\dagger$} Jean-Philippe Bourgoin,\authormark{1} Ramy Tannous,\authormark{1} Sascha Agne,\authormark{1} Christopher J. Pugh,\authormark{1} Katanya B. Kuntz,\authormark{1} Brendon L. Higgins\authormark{1} and Thomas Jennewein\authormark{1,2$\ddagger$}}

\address{\authormark{1}Institute for Quantum Computing and Department of Physics and Astronomy, University of Waterloo, Waterloo, ON, N2L 3G1 Canada\\
\authormark{2}Quantum Information Science Program, Canadian Institute for Advanced Research, Toronto, ON, M5G 1Z8 Canada \\
\authormark{$\dagger$} jeongwan.jin@nrc-cnrc.gc.ca\\
\authormark{$\ddagger$} thomas.jennewein@uwaterloo.ca\\
 }



\begin{abstract}
Despite its widespread use in fiber optics, encoding quantum information in photonic time-bin states is usually considered impractical for free-space quantum communication as turbulence-induced spatial distortion impedes the analysis of time-bin states at the receiver. Here, we demonstrate quantum key distribution using time-bin photonic states distorted by turbulence and depolarization during free-space transmission. Utilizing a novel analyzer apparatus, we observe stable quantum bit error ratios of 5.32\,\%, suitable for generating secure keys, despite significant wavefront distortions and polarization fluctuations across a 1.2\,km channel. This shows the viability of time-bin quantum communication over long-distance free-space channels, which will simplify direct fiber/free-space interfaces and enable new approaches for practical free-space quantum communication over multi-mode, turbulent, or depolarizing channels.
\end{abstract}

\section{Introduction}
 Networks of quantum information processors\textemdash a Quantum Internet\textemdash demand the reliable transmission of quantum bits between nodes via optical-fiber or free-space channels\,\cite{Kimble:2008}. Encoding quantum information in photonic time-bin states is often preferred for realizing quantum-network components in optical fibers; e.g., boson sampling\,\cite{He:2017}, optical-fiber quantum memory\,\cite{Saglamyurek:2011}, city-wide quantum teleportation\,\cite{Valivarthi:2016}, and high-dimensional quantum key distribution (QKD)\,\cite{Islam:2017}. However, time-bin encoding is usually considered impractical in free space as conventional analyzers are hindered by the effects of atmospheric turbulence and steering-optics misalignment\,\cite{Jin:2018}. In fact, free-space quantum communication has been demonstrated by utilizing different degrees of freedom on various platforms, including polarization-encoded satellite QKD\,\cite{Liao:2017}, orbital-angular-momentum-encoded high-dimensional QKD\,\cite{Sit:2017}, and daylight QKD with integrated chips\,\cite{Avesani:2019}. Hyper-entanglement between energy-time and polarization has also been distributed across 1.2\,km free-space link\,\cite{Steinlechner:2017}.
 
Practical time-bin encodings of quantum bits (qubits) must have a sufficiently large temporal separation between \emph{early} and \emph{late} time-bins that optical pulses are easily distinguishable in the presence of system timing jitters---e.g., on the order of 1\,ns for direct detection. This temporal separation defines the difference in lengths of the two paths of an asymmetric interferometer used to analyze the time-bin states. But, when given an optical mode that is spatially distorted by atmospheric 
turbulence, the evolutions of the spatial modes through each of those two paths will be significantly distinguishable\,\cite{Jin:2018}. Consequently, the interference contrast of superpositions of early and late time-bins is degraded. Spatial-mode filtering via coupling to single-mode fiber can be used to avoid this problem, but doing so results in significant throughput loss for long distance links\,\cite{Takenaka:2012}. Wavefront corrections with adaptive optics can help, but is expensive and technically challenging\,\cite{Liu:2019,Chen:2018}. Therefore, a purely passive multimodal optical solution is preferable.

Recent work has achieved significant progress on passive optical solutions. For instance, Jin \textit{et al} showed two approaches of field-widened interferometers used as analyzers for a multi-mode photonic time-bin qubit entangled with a separate polarization qubit, proving the viability of time-bin encoding for quantum communication in multi-mode optical channels\,\cite{Jin:2018}. A similar concept was implemented by Vallone {\it et al}, who observed interference between two temporal modes of laser pulses which were attenuated to single-photon levels after being reflected by orbiting satellites\,\cite{Vallone:2016}. Until now, however, a genuine time-bin-based quantum protocol over a long-distance atmospheric link has not been demonstrated. 

Here, we demonstrate QKD between two separate parties using time-bin encoding as an alternative means for free-space quantum communication. First, we prepare time-bin quantum states, encoded in early and late time-bin with a 2-ns separation, and send them through a polarization-mixing channel before traveling 1.2\,km of free-space atmosphere and becoming spatially distorted by turbulence. At the receiver, the polarization- and spatially-distorted photons are coupled into a step-index multi-mode fiber, which supports approximately 1870 propagating spatial modes\,\cite{Papadopoulos:2012}. The collected photons are directed to our multi-mode time-bin qubit decoder (MM-TQD) where passive mode-correcting optics ensure high quality of interference regardless of the spatial shape of the incoming photons.

\section{Experiment} 
\begin{figure*}[htpb]
\centering\includegraphics[width=13.5cm]{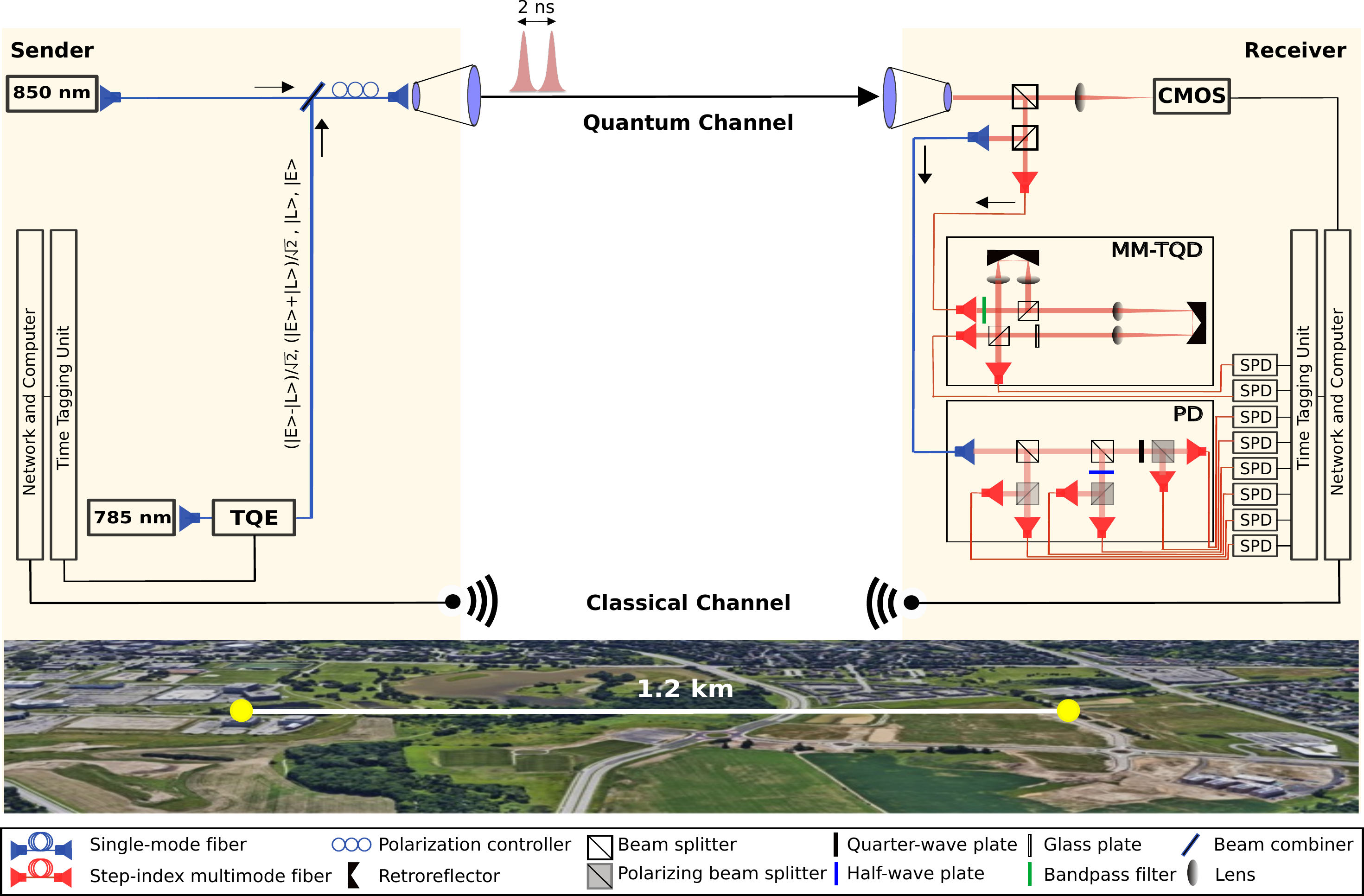}
\caption{ {\bf Experimental setup.} Time-bin photonic qubit states,  $|\rm{E}\rangle$, $|\rm{L}\rangle$, $(|\rm{E}\rangle+|\rm{L}\rangle)/ \sqrt{2}$, or $(|\rm{E}\rangle-|\rm{L}\rangle)/ \sqrt{2}$, are generated in a time-bin qubit encoder (TQE). The signals traverse 1.2\,km of atmosphere to the receiver, and are coupled via multi-mode fiber to time-bin qubit decoder (MM-TQD) for analysis. A separate laser beam at 850\,nm co-propagates with time-bin-encoded photons, and is used to characterize the degree of turbulence and polarization mixture in the optical channel. A spatial image of this light is captured by the CMOS sensor, and its polarization state is analyzed by the 6-state polarization decoder (PD). All detected signals from single-photon detectors (SPDs) are sent to a time-tagging unit and a computer for data analysis. Classical communication is done through a radio-frequency local network channel.}
\label{setup}
\end{figure*}

We implement the BB84 QKD protocol\,\cite{Bennett:1984} with decoy states\,\cite{Hwang:2003, Wang:2005, Ma:2005}. Our apparatus (Fig.~\ref{setup}) prepares time-bin qubits using 300-ps-long 785-nm-wavelength weak coherent pulses, with ${\approx}0.5$ mean photons per pulse, generated at a rate of 150\,MHz. Although other wavelengths may be used, 785\,nm is an excellent compromise for satellite-uplink long-distance QKD\,\cite{Bourgoin:2013}, while portable and room-temperature-operating single-photon detectors provide high detection efficiency. For information encoding, we use a time-bin qubit encoder (TQE) consisting of an unbalanced interferometer with path length difference matched to the MM-TQD. Where $|\rm{E}\rangle$ and $|\rm{L}\rangle$ represent early and late time-bins, respectively, the state $|\rm{E}\rangle$, $|\rm{L}\rangle$, $(|\rm{E}\rangle+|\rm{L}\rangle)/ \sqrt{2}$, or $(|\rm{E}\rangle-|\rm{L}\rangle)/ \sqrt{2}$ is encoded onto each pulse, following a randomized sequence. (See Appendix A for details.) The time-bin-encoded photon is then coupled into a single-mode fiber and guided to a 12-cm-diameter launch telescope. 

After traversing the free-space channel located at the North Campus of the University of Waterloo, the single photons are captured by a 10-cm-diameter receiver telescope and collected in a multi-mode fiber which guides the light to our MM-TQD for projection measurements. The MM-TQD consists of an unbalanced interferometer with a matched pair of lenses in each path, minimizing relative dispersion. The focal lengths of each lens pair, $f_{\rm{short\,path}}=50\,\text{mm}$ and $f_{\rm{long\,path}}=200\,\text{mm}$, are chosen such that the image of the incoming spatial mode at the input beam splitter is recovered at the output beam splitter after traversing each path of the interferometer, thus removing path distinguishability caused by path-length differences while maintaining the temporal separation between the paths. For an input multi-mode beam, we measure interference visibility of up to 97\,\%. Photons are collected into multi-mode fiber, with total throughput of 81\,\% from input to output, and are finally detected by silicon avalanche single-photon detectors (Si-SPDs). A QKD post-processing algorithm is applied to the data, including error correction and privacy amplification (see Appendix B), to produce secret key bits.

To characterize the channel, a separate ${\sim}100\,\mu\text{W}$ 850-nm-wavelength laser is polarized and merged with the quantum signals via a single-mode beam combiner at the transmitter before entering the launch telescope. At the receiver, a portion of this light is captured by a complementary metal-oxide semiconductor (CMOS) sensor at a frame rate of 20\,Hz for 60\,s at a time, by which we assess beam distortions and spatial misalignment caused by the turbulent atmosphere. A portion of the laser beam is coupled into a single-mode fiber before undergoing tomographic polarization analysis\,\cite{Altepeter:2005}---consisting of 6-state projection and photon counting with Si-SPDs (see Appendix C)---to assess the polarization preservation of the channel. A bandpass filter centered at 785\,nm is placed at the entrance of the MM-TQD to block 850-nm laser light from the time-bin-state measurements. Intentionally blocking the transmission channel for brief periods during data collection allows us to synchronize these additional measurements with the QKD data.

Thermal and mechanical instabilities cause phase shifts between the TQE and MM-TQD, which we compensate with a motorized glass plate located in the long path of the TQE, manually optimizing for minimal QBER. An identical glass plate is placed in the long path of the MM-TQD to compensate glass-induced dispersion between the interferometers. Although channel delay fluctuations induced by atmospheric turbulence\,\cite{Kral:2005} are below our time-bin resolution, note also that the typical time scales of turbulence-induced phase shifts, on the order of 10 to 100 ms\,\cite{Gisin:2002}, significantly exceeds the temporal separation between our time-bins. Thus, the temporal delay each time-bin experiences during the passage of turbulence is effectively equal (as evidenced by the low $\text{QBER}_\text{time}$ values measured).

\section{Results} 

\begin{figure*}[htpb]
\centering\includegraphics[width=9.5cm]{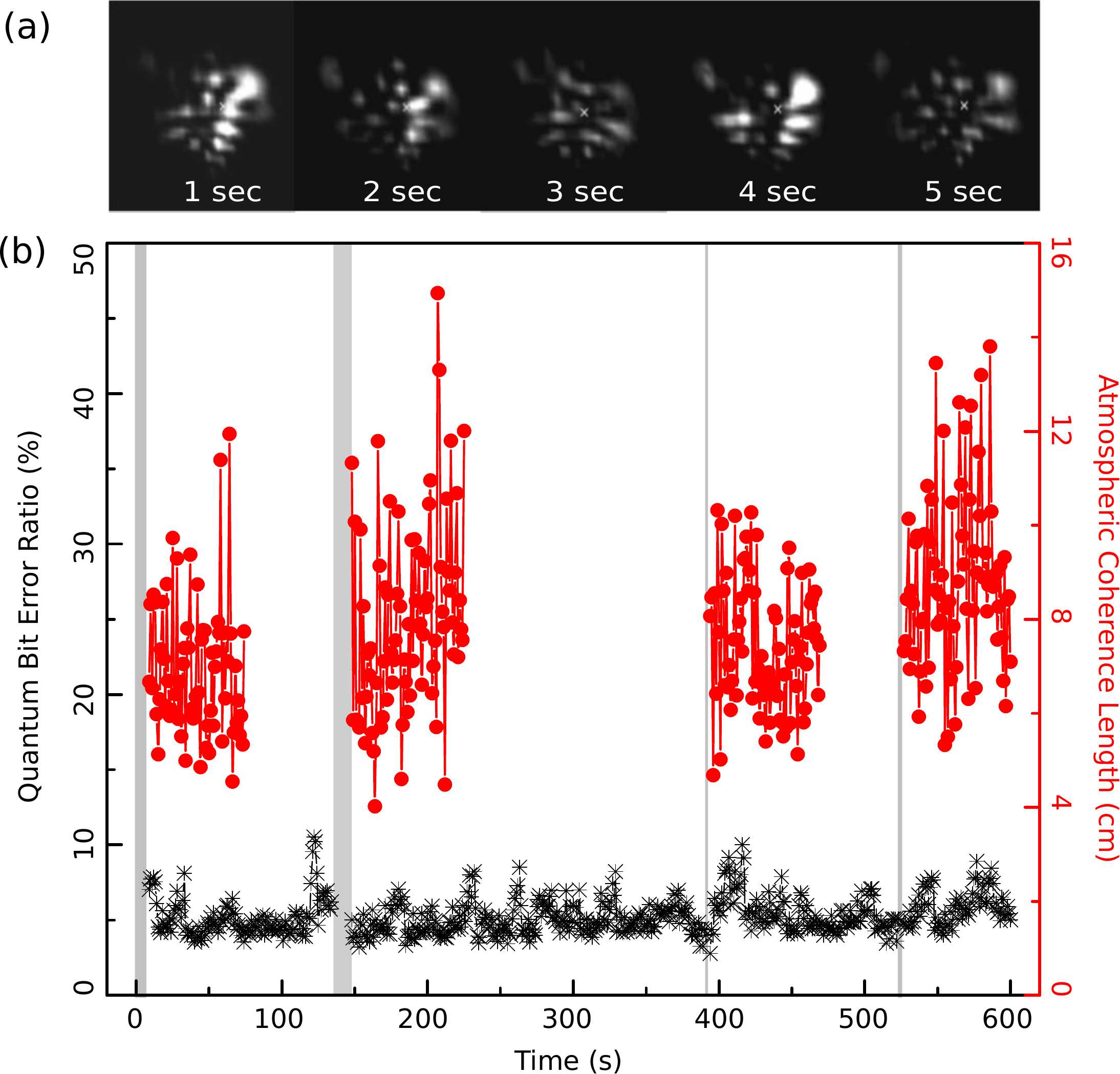}
\caption{ {\bf Characterization of the turbulent quantum channel of 1.2\,km.} (a) Selected spatial-mode snapshots of incoming laser light at the receiver, captured by the CMOS sensor. A constant background pattern, caused by optical reflections, is significantly modulated by turbulence-induced distortion. (b) Despite short and fluctuating atmospheric coherence (red dots) of the optical link, low and steady values of $\rm{QBER}_{time}$ (black asterisks) are observed. On occasion (highlighted by the grey areas), signals are intentionally blocked for synchronization.}
\label{turbulence}
\end{figure*}

We first demonstrate the robustness of our system against atmospheric turbulence. To assess the quality of optical transmission through a turbulent channel, we characterize the channel's atmospheric coherence length, or Fried parameter\,\cite{Fried:1966}, $r_0$. This parameter is inversely proportional to turbulence strength, and can quantify the cumulative turbulence effect over the entire free-space link. We estimate $r_0$ by analyzing the beam centroid motion through sequential CMOS sensor frames (see Fig.~\ref{turbulence}(a) and Appendix D). The measured atmospheric coherence lengths, averaged per second, fluctuate between 6 to 10\,cm over the duration of our 600\,s measurement, as shown in Fig.~\ref{turbulence}(b), with a mean value of $r_0^\text{mean} = \text{7.83\,cm}$. For comparison, this is equivalent to a refractive-index-structure parameter $C_n^2$ of $2.78 \times 10^{-15}\,\text{m}^{-2/3}$.  In a much less-turbulent channel with $C_n^2$ of $3.00 \times 10^{-17}\,\text{m}^{-2/3}$, Schmitt-Manderbach \textit{et al} generated 28~bits/s of QKD key using a polarization encoding and a pulse rate of 10\,MHz\,\cite{Schmitt-Manderbach:2007} (which would scale to 420~bits/s at 150\,MHz repetition), while a $C_n^2$ of around $1.70 \times 10^{-14}\,\text{m}^{-2/3}$ is considered somewhat typical at sea-level, in the context of nighttime astronomical observations\,\cite{Tofsted:2006}. Despite the fluctuating coherence of the atmosphere, we observe relatively stable $\rm{QBER}_{time}$ with a mean value of 5.32\,\%. Note that about 2 percentage points of $\rm{QBER}_{time}$ stem from time-bin qubit source preparation, while synchronization blockages cause transient QBER spikes to around 50\,\%.
We employ QKD post-processing of the data, including signal-to-noise filtering which suppresses contributions from these blockages, and generate secret keys at a rate of 154.2~bits/s in the asymptotic limit, under a total system loss of 38.4\,dB (see Table~\ref{qkd-parameters} and Appendix B). In comparison, the visibility $V=\text{16\,\%}$ achievable at the decoder without lenses\,\cite{Jin:2018}, with $\rm{QBER} \geq (1-V)/2$\,\cite{Gisin:2002}, equates to a QBER of at least 42\,\%, well in excess of the 11\,\% bound for qubit-QKD protocols to succeed, even in principle.

\begin{figure*}[htpb]
\centering\includegraphics[width=12cm]{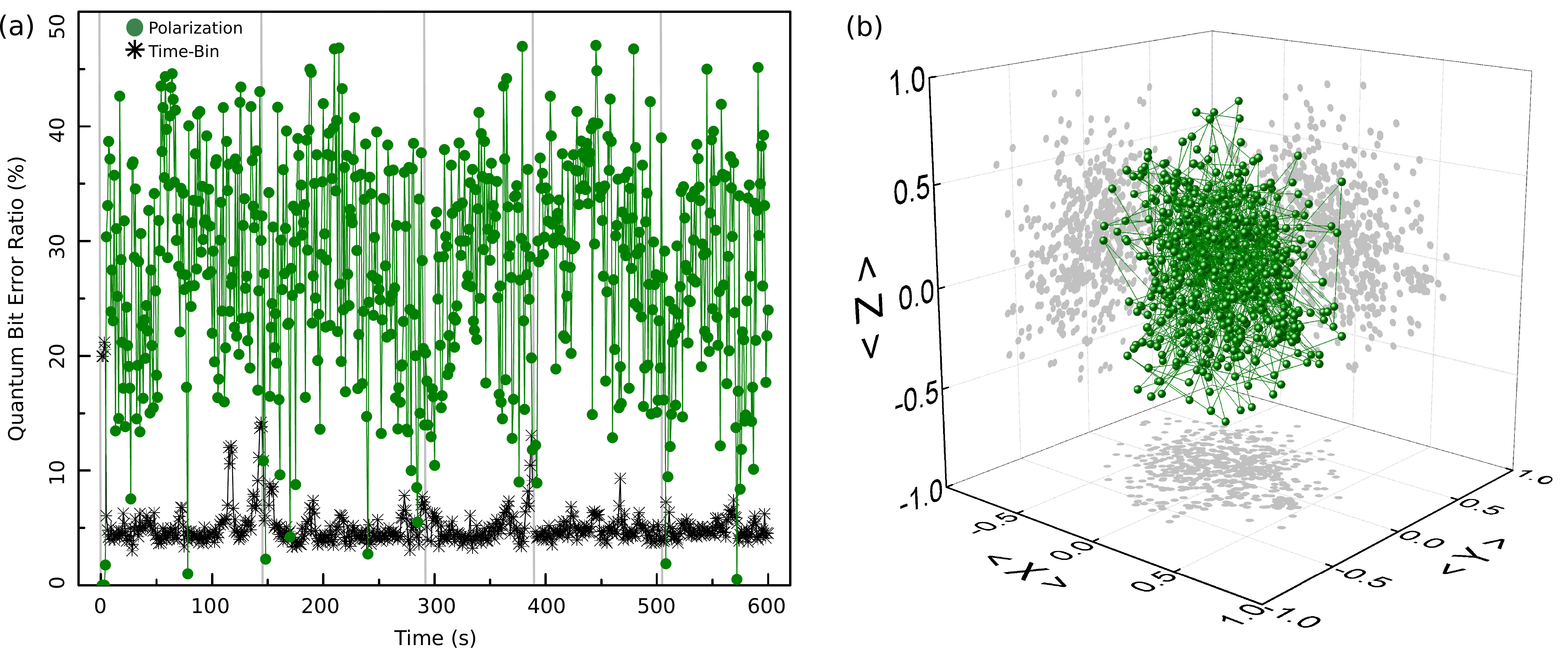}
\caption{ {\bf Characterization of the depolarizing and turbulent quantum channel.} (a) Green dots represent effective achievable $\rm{QBER}_{pol}$ of polarization states through the channel, estimated from reconstructed density matrices. The measured stable $\rm{QBER}_{time}$ (black asterisks) confirms robustness of the time-bin encoding against polarization mixture. Grey areas indicate when signals are blocked for synchronization. (b) Distribution of measured polarization expectation values.}
\label{depolarization}
\end{figure*}

Next, we demonstrate our time-bin QKD system's robustness against depolarizing effects. To achieve these, we mix the polarization in the channel by continuously manipulating (by hand) the single-mode fiber that carries time-bin-encoded photons to the transmitter, i.e. rapid transforms of polarization. Then, for each density matrix $\rho$ produced after 1\,s of photon counting by the tomographic polarization analyzer at the receiver, we calculate the qubit purity of the measured polarization, $P = \operatorname{Tr}(\rho^2)$. The effective achievable QBER for the polarization encoding, $\rm{QBER}_{pol}$, is estimated via relating the purity $P$ to QBER via the visibility $V=\sqrt{2P-1}$\,\cite{Bussieres:2014} (assuming rotationally symmetric noise). Such analysis then corresponds to finding the least QBER that could possibly be achieved by a polarization correction mechanism operating at a 1\,Hz bandwidth.

Results are shown in Fig.~\ref{depolarization}. Projections of $\rho$ along each of the Pauli bases (Fig.~\ref{depolarization}(b)) are spread over one large cluster, indicating significant variation of the polarization states. Over time the corresponding polarization state QBER values (Fig.~\ref{depolarization}(a)) fluctuate between almost pure ($\rm{QBER}_{pol} \approx 0\,\%$) and almost completely mixed ($\rm{QBER}_{pol} \approx 50\,\%$). Simultaneously, we observe steady time-bin QBER around the 5.08\,\% mean, with no correlation to polarization QBER. After post-processing of time-bin measurements, we find secure keys generated at a rate of 138.8\,bits/s in the asymptotic regime, while polarization did not permit secure key generation. 

\section{Conclusion}
We experimentally demonstrated QKD using time-bin encoded photons in a free-space channel of 1.2 km range, under distortions of turbulence as well as turbulence \& depolarization. Despite turbulence-induced wavefront distortions on the transmitted photons, our passive multi-mode time-bin qubit decoder remains efficient and of high interference-visibility, allowing the system to achieve a stable mean time-bin quantum bit error ratio of 5.32\,\%. This channel is capable of generating secure keys at a rate of 154.2\,bits/s in the asymptotic limit. Moreover, in the presence of rapid polarization changes the error ratios remain essentially unchanged around 5.08\%, with 138.8\,bits/s asymptotic secure key rate. Our system could further be improved by miniaturzing the time-bin analyzer for better thermal stability, or by automating phase compensation. (Implementation of a reference-frame-independent protocol could even relax stability requirements of the time-bin interferometers.) Future study may also include development of a highly thermally stable unbalanced interferometer for analysis of time-bin states using different refractive index materials in each path, which could also allow extension into analysis of higher-dimensional states. By enabling quantum communication with time-bin encoding over multi-mode or depolarizing free-space channels, our demonstration opens possibilities to translate advanced fiber-based quantum protocols, such as coherent-one-way and differential-phase-shift quantum key distribution protocols, to long-distance free-space channels. Furthermore, encoding technologies used in quantum networks can be simplified by directly interfacing optical-fiber and free-space quantum links using time-bin-encoded photons.

\begin{table}[tbp]
 \centering \caption{ {\bf Measured QKD parameters, taken between 3 and 4\,AM local time on August 17th, 2017.} Low values of $\rm{QBER}_{time}$ allow the creation of secret keys in the asymptotic regime.}
\label{qkd-parameters}
\centering \begin{tabular}{ccc}
\hline
& Turbulent & Turbulent \& Depolarizing \\
    \hline
    Signal photons/ pulse & 0.488 & 0.520 \\
    \hline
    Decoy photons/ pulse & 0.082 & 0.094 \\
    \hline
    Background/ pulse & $3.65\times 10^{-7}$ & $3.45\times 10^{-7}$ \\
    \hline
    Channel loss (dB) & 38.4 & 38.8  \\
    \hline
$\rm{QBER}^{mean}_{time}$ (\%) & 5.32 & 5.08 \\
\hline
Key rate (bits/s)  & 154.2 & 138.8 \\
    \hline
   \end{tabular}
    \end{table}
\section*{Appendix A: Time-bin qubit encoder}
We chose to utilize an available existing polarization-encoding source for convenience. (A direct time-bin encoding might reduce the overall complexity of the apparatus while increasing fidelity.) This source generates 785-nm-wavelength weak coherent pulses at 150\,MHz that are polarization-encoded using a pair of electro-optical phase modulators in a balanced Mach-Zehnder interferometer\,\cite{Pugh:2017}. Each pulse is initialized in one of four BB84 polarization states: horizontal ($|\rm{H}\rangle$), vertical ($|\rm{V}\rangle$), diagonal ($|\text{D}\rangle = (|\rm{H}\rangle+|\rm{V}\rangle)/ \sqrt{2}$), or antidiagonal ($|\text{A}\rangle = (|\rm{H}\rangle-|\rm{V}\rangle)/ \sqrt{2}$). The state is selected for each pulse following a randomized repeating sequence. (Though not secure, this does not invalidate the proof-of-principle conclusions we draw, and can be corrected with appropriate electronic engineering.) Signal and decoy intensity levels are controlled by a separate electro-optical intensity modulator, producing mean photon numbers per pulse of ${\approx}0.5$ and ${\approx}0.1$, respectively. These are measured at the transmitter with a calibrated pick-off and single-photon detector. Vacuum states are achieved by suppressing the laser trigger. Each intensity state appears in the sequence in proportions of 80\,\% (signal), 14\,\% (decoy), and 6\,\% (vacuum). After propagating through single-mode optical fiber, these pulses are then coherently converted to time-bin qubit states by an unbalanced polarizing Michelson interferometer~\cite{Jin:2018}, with path length difference matched to the MM-TQD. Polarization drifts in the fiber are corrected by a system that performs once-per-second tomographic reconstructions of a pick-off, and compensates using a quarter-, half-, quarter-wave plate triplet with optimized orientations~\cite{Higgins:2018}.

\begin{figure*}[htpb]
\centering\includegraphics[width=9.5cm]{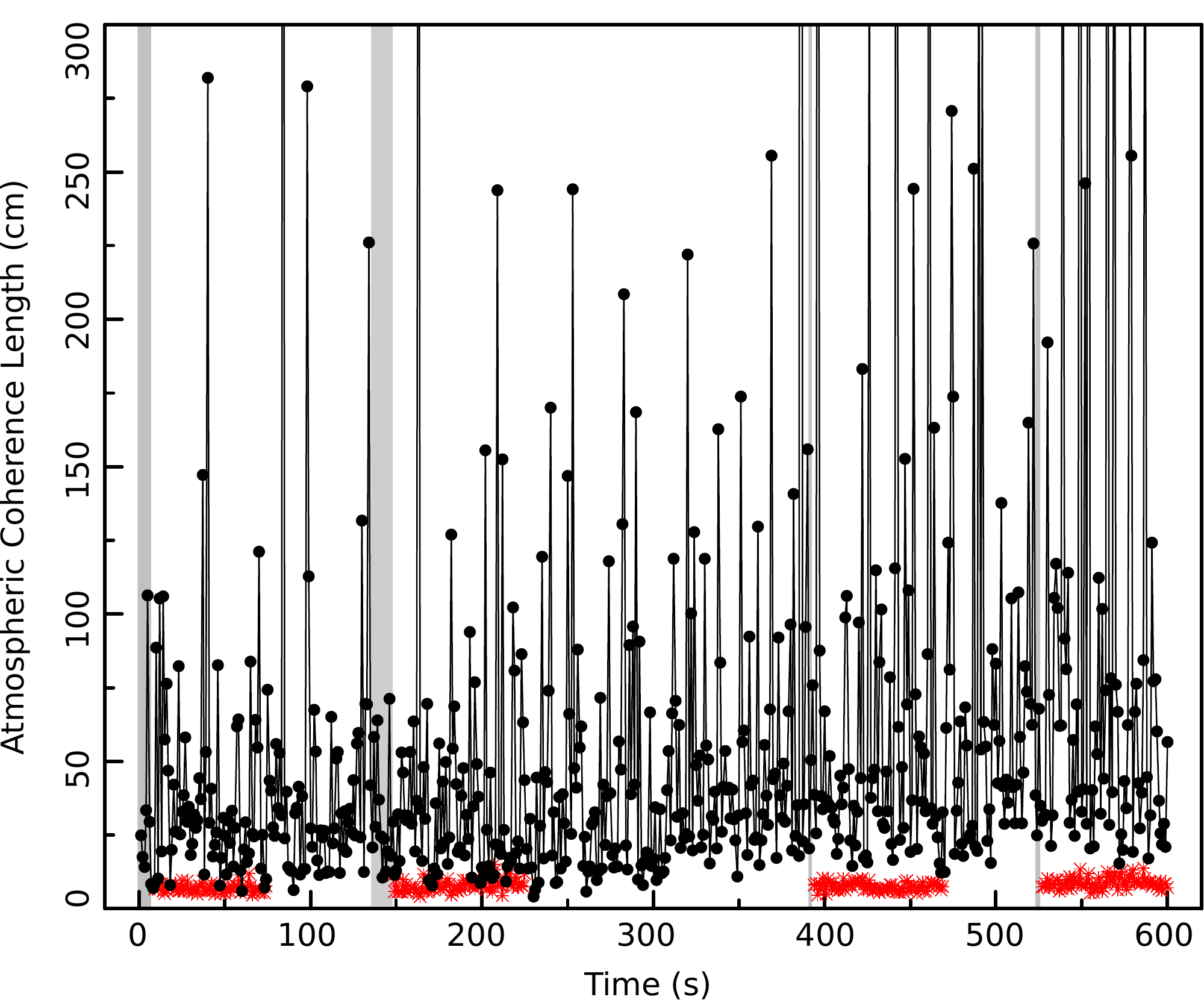}
\caption{ {\bf Turbulence strength comparison.} Measured coherence lengths of the outdoor channel (red asterisks) are much shorter than indoor values (black circles), showing that the spatial coherence of a photon will not be as well preserved over the longer outdoor transmission. The image recording rate for the outdoor (indoor) measurement is 20\,Hz (2\,Hz). Interruptions during the outdoor experiment are due to limited camera memory capacity. Upon resetting the camera, we intentionally blocked the beam (indicated by grey areas) to synchronize with single-photon detections.}
\label{coherence}
\end{figure*}

\section*{Appendix B: Postprocessing for secret key generation}
Due to restrictions of the available software, time correlation processing is performed at the transmitter---this is sufficient for our proof-of-principle demonstration, however a real-world implementation would have this analysis performed at the receiver or otherwise done in a way that suppresses the timing side-channel vulnerability\,\cite{Lamas-Linares:2007}. Time correlation is done on per-second-aggregated data, with delays optimized on the number of source emission events coincident with reception events in the time-bin superposition basis. Early (optimized delay plus 2\,ns) and late (optimized delay minus 2\,ns) coincidences are then also identified. We use a signal-to-noise filter to discard per-second regions with low count rate (under 500\,Hz), such as times when the signal is blocked to synchronize data across QKD, spatial-mode, and polarization measurements (these blockages are not included in the 600\,s total measurement time). Such regions are treated as if no pulses were emitted from the source or detected by the receiver---otherwise, these sections would worsen the QBER due to the background noise\,\cite{Erven:2012}. The remaining data are basis-sifted to produce the sifted key. Error correction is accomplished using low-density parity-check codes: the receiver calculates an error correction syndrome from its sifted key based on a matrix constructed by the transmitter, which then uses this syndrome to reconcile its sifted key with the receiver's (observed error correction efficiency around 1.17). The resulting error-corrected keys are privacy amplified using an augmented Toeplitz matrix acting as a two-universal hash\,\cite{Bourgoin:2015}, producing the final secret key. We do not consider the effects of finite-size statistics, but calculate the length of the final key assuming the measured values extrapolate to the asymptotic limit.

\section*{Appendix C: Polarization preservation characterization}
At the receiver, the 850-nm-wavelength light is analyzed in a tomographically complete basis set, consisting of horizontal/vertical, diagonal/antidiagonal, and right-circular/left-circular polarization projections by an arrangement of wave plates, beam splitters, and polarizing beam splitters (see Fig.~\ref{setup}). Photons are then coupled to multi-mode fibers and detected by Si-SPDs. For each 1\,s integration of photon detection counts, maximum likelihood estimation\,\cite{Altepeter:2005} then reconstructs the density matrices of the polarization states. A completely pure state of polarization, within tomographic uncertainty, is measured when the single-mode fiber is untouched at the sender. Note that the polarization state undergoes a unitary rotation in the single-mode fiber that takes photons from the collection telescope to the tomographic analyzer. As the fiber is stationary, this unitary varies slowly over time (primarily due to temperature fluctuation), well outside the 1\,s analysis time-scale. Because of this, the measured purity of the polarization state---and thus, the estimated achievable $\text{QBER}_\text{pol}$---is accurate even though the actual rotation is not known.

\section*{Appendix D: Atmospheric coherence length}
The atmospheric coherence length $r_0$ is directly related to the tilt angle variance of centroid displacement for two uncorrelated axes, $\sigma^2_\text{2-axis}$, as\,\cite{Tyson:2010}
\begin{equation} \label{eq:alpha}
\sigma^2_\text{2-axis} = 0.364\bigg(\frac{D \lambda}{r_0^2} \bigg)^{5/3},
\end{equation}
where $D=\text{12\,cm}$ is the initial beam diameter and $\lambda = \text{850\,nm}$ is its wavelength. We first subtract background noise from each CMOS sensor frame capture of the beam that may bias the center-of-mass calculation. The sensor frame rate for the data shown in Fig.~\ref{turbulence}(b) was 20\,Hz. Taking the centroid variance from the mean over 20 frames leads to a mean $r_0$ for each second of measurement.

\begin{figure*}[htpb]
\centering\includegraphics[width=9.5cm]{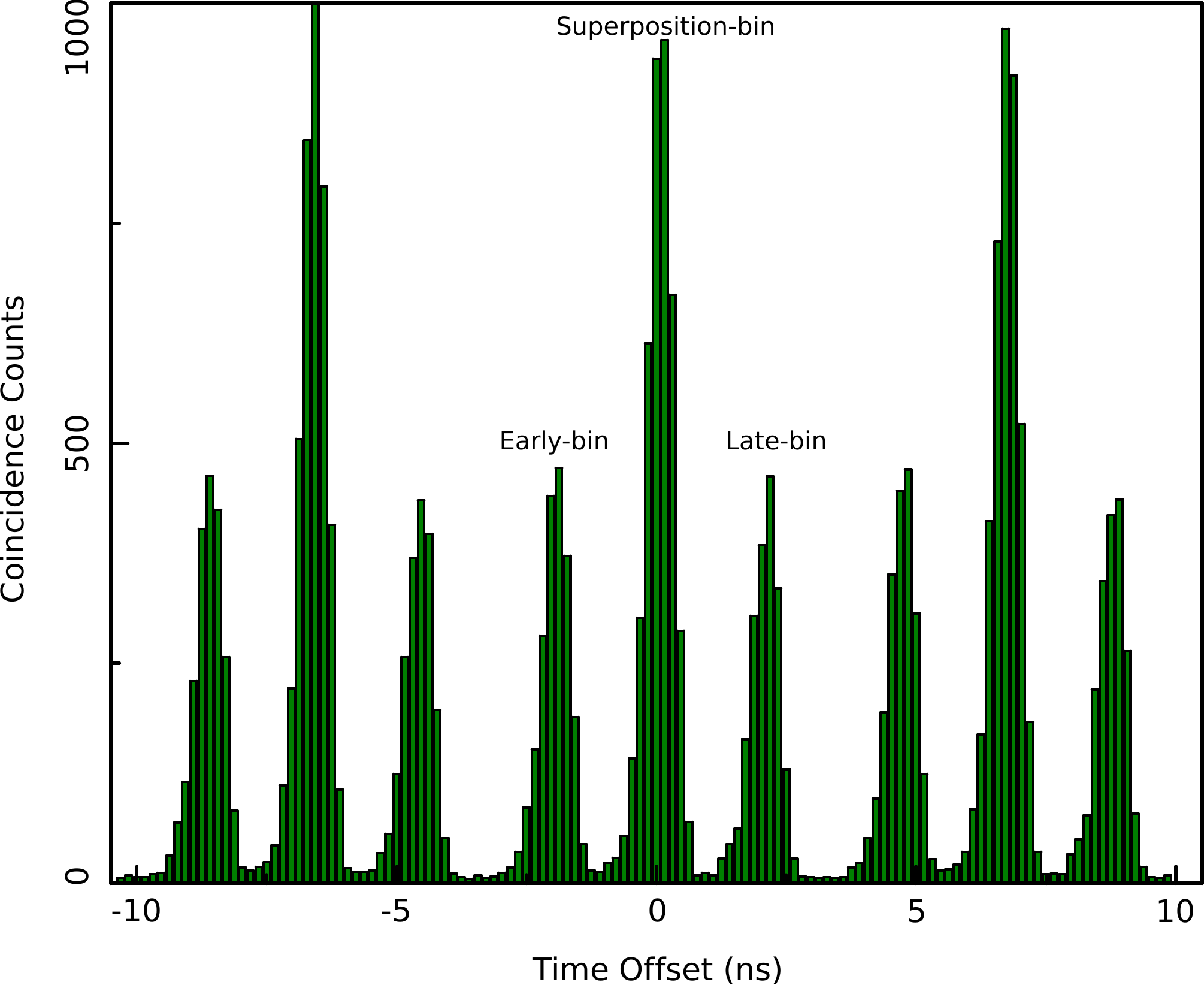}
\caption{ {\bf  Typical timing histogram.} Shown are the counts of coincidences of source emissions with detection events, given a varying time offset between events at source and receiver (after correcting for delays, including the $4\,\mu\text{s}$ free-space time-of-flight) and a 1\,ns window, for one second of outdoor free-space link data. The time bins of the encoded quantum signal are clearly visible, with the central peak (around time offset 0) corresponding to the superposition basis, and the first minor peaks on either side corresponding to early and late states being measured. Other peaks originate from adjacent pulses, owing to the periodic pulsing of the weak-coherent pulse source, and are not coincidences of related events.}
\label{histogram}
\end{figure*}

\section*{Appendix E: Turbulence strength}
The coherence lengths of an indoor atmosphere are measured for comparison to those of the outdoor free-space channel. For this we capture the spatial images of classical light at 850\,nm, co-propagated with time-bin quantum signal, using a CCD camera at a rate of 2\,Hz. From each captured frame a centroid value is calculated, and from the variations of these we calculate mean coherence lengths per second using Eq.~1. As shown in Fig.~\ref{coherence}, the measured indoor values range from a few centimeters to several meters, yielding an average value of $l_\text{in}^\text{avg} = 65.8\,\text{cm}$, which significantly exceeds the measured outdoor values of $l_\text{out}^\text{avg} = 7.83\,\text{cm}$. In order to assess the degree of $r_0$ fluctuation, we compute its standard deviation of $\sigma_{\text{indoor}} = 65.3\,\%$ and $\sigma_{\text{outdoor}} = 25.4\,\%$ (relative to their respective means) using data averaged over the same size of samples. We believe that the bigger variation in indoor measurement is caused by large and rapid temperature changes over a short part of the beam path, mainly owing to the building's heating and ventilation systems, which were co-located with the apparatus. The fact that the outdoor coherence length is much shorter than the indoor value despite slower temperature changes is because the heat fluctuation occurs over much longer distance.

\section*{Appendix F: Resolution of time bins in coincidence analysis}
Fig.~\ref{histogram} shows a typical time-correlation histogram for one second duration of measured detection events of transmission over the 1.2\,km free-space channel. The histogram counts correspond to the number of detection events coincident with source emission events given a delta time offset. Early, late, and superposition time bins are clearly resolvable. The width of the time bins is dominated by the timing jitter of our single-photon detector, which is about 500\,ps at the full width at half maximum. Clear separation between time bins, larger than timing jitter induced by any involved devices, is critical for communication.

\section*{Funding}

The Office of Naval Research; The Canadian Space Agency; The Canada Foundation for Innovation (25403, 30833); The Ontario Research Fund (098, RE08-051); The Canadian Institute for Advanced Research; The Natural Sciences and Engineering Research Council of Canada (RGPIN-386329-2010); Industry Canada.

\section*{Disclosures}
The authors declare no conflicts of interest.


\bibliography{TBQKD}






\end{document}